\documentstyle[PASJadd,psbox,epsf]{PASJ95}
\newcommand{\lett}{}

\begin{document}

\title{ASCA Observations of the Twin Supernova Remnants in the Large
Magellanic Cloud, DEM L316}

\author{Mamiko {\sc Nishiuchi}, Jun {\sc Yokogawa}
 and Katsuji {\sc Koyama}\thanks{CREST, Japan Science and 
Technology Corporation (JST), 
  4-1-8 Honmachi, Kawaguchi, Saitama, 332-0012} \\
   {\it Department of Physics, Graduate School of Science, Kyoto University, 
   Sakyo-ku, Kyoto, 606-8502} \\
   {\it E-mail(MN): mamiko@cr.scphys.kyoto-u.ac.jp}\\
and\\
John P. {\sc Hughes}\thanks{ also Service d'Astrophysique, L'Orme des Merisiers, CEA Saclay, 
91191 Gif-sur-Yvette Cedex FRANCE} \\
{\it Department of Physics and Astronomy, Rutgers University, 136 Frelinghuysen Road, 
Piscataway, NJ 08854-8019}\\}

\abst{ We report results from an ASCA X-ray study of DEM L316, an
emission nebula in the Large Magellanic Cloud (LMC) consisting of two
closely-spaced supernova remnants (SNRs).  The SIS image shows
separate X-ray sources located at the positions of the two radio- and
optically-emitting SNR shells, 0547$-$69.7A and 0547$-$69.7B
(hereafter, shell A and B).  The individual X-ray spectrum of each
shell is well described by optically-thin thermal emission, 
although the characteristics of
the emission differ in important details between them. Shell A
exhibits strong iron L emission, which we attribute to the presence of
iron-rich ejecta leading to the suggestion that this SNR originates
from a Type Ia SN, an explosion of a moderate mass progenitor.  Shell
B, on the other hand, has a chemical composition similar to that of
the interstellar medium of the LMC and so its X-ray emission is
dominated by swept-up matter.
The different spectral features
strongly disfavor the hypothesis that the two shells are due to one
supernova explosion into an interconnected bubble.  
We could not obtain the evidence which
positively supports the collision between two SNRs.  }

\kword{SNR: individual (DEM L316) ---
galaxies: individual (LMC) --- interstellar matter --- X-rays: SNRs}

\maketitle
\thispagestyle{headings}

 \section{Introduction}

Nuclear burning in stars and their subsequent explosions as supernovae
(SNe) are the primary means by which heavy elements are
synthesized and distributed in the interstellar medium (ISM) of
galaxies.  The chemical composition of SN ejecta depends strongly on
the type of the SN explosion and the mass of the progenitor.  For
example, type Ia SNe, which produce iron-rich ejecta, are believed to
be the incineration of a CO white dwarf driven to the Chandrasekar
limit by accretion (Nomoto, Thielemann, \& Yokoi 1984).  
A type II (or Ib/c) SN, the core-collapse of a
massive star, tends to eject large amounts of oxygen produced during
the hydrostatic evolution of the progenitor (Tsujimoto et al. 1995). 
The mix of SNe types,
plus the nucleosynthetic yield of individual SNe, is an essential
element that determines the chemical evolution of galaxies and the
structure and dynamics of the ISM.  SNe often produce remnants (SNRs)
containing hot ($10^7$ K) X-ray emitting plasma with spectra
characterized by numerous emission lines from heavy elements.  X-ray
spectroscopy therefore provides a powerful tool for determining the
chemical composition of SN ejecta and the ambient medium, as well as
the amount of energy input to the ISM.

The radio SNR DEM L316 is located in the nearest sizeable galaxy, the
Large Magellanic Cloud (LMC), at a distance of $\sim$ 50 kpc (van den
Bergh 2000).  The LMC's nearly face-on geometry, or small inclination
angle (30$^{\circ}$ -- 40$^{\circ}$; Feast 1991), and its location at
high Galactic latitude result in only modest interstellar absorption
for individual objects in the Cloud.  Thus, unlike most Galactic SNRs,
DEM L316 has a reliable distance estimate and low ISM absorption.  DEM
L316 consists of a pair of closely-spaced, limb-brightened shells of
radio and optical emission, called SNR 0547$-$69.7A and SNR
0547$-$69.7B (hereafter shell A and shell B).  It has been proposed
that these twin shells are (1) two independent remnants superposed
along the line of sight (Mathewson and Clarke 1973), (2) a single SN
that exploded into an interconnected bubble formed by a stellar wind
or a previous supernova (Lasker 1981; Mills et al.~1984), or (3)
``colliding remnants'' (Williams et al.~1997).  Because of the
similarity in their mean radial velocities, some authors (Lasker 1981;
Williams et al.~1997) have argued that the two shells are physically
co-located.  In fact the individual mean radial velocities of the two
shells are only poorly constrained, largely because they each come
from averaging over the complex velocity field of an incomplete SNR
shell expanding at a velocity of $\sim$200 km s$^{-1}$ (Williams et
al.~1997).  
Unfortunately Williams et al.~(1997) do not 
assess quantitatively
the level of difference in mean velocity of the two shells allowed by
their data.  
These authors also found enhanced [O {\sc iii}] emission
and a change in the magnetic field structure in the region between
shell A and B, from which they inferred that the two shells in DEM
L316 were indeed colliding SNRs.

Our recent study of LMC SNRs (Hughes, Hayashi, \& Koyama 1998) has
shown that ASCA X-ray spectra can provide powerful constraints on the
nature, evolution, and composition of remnants.  As part of our
continuing research effort in this area we undertook ASCA observations
of DEM L316.  The current article presents our detailed X-ray spectral
study of each shell using nonequilibrium ionization models.  We find
that the ASCA data do indeed provide important insights into the
nature of DEM L316 and shed light on the relationship between 
shells A and B.

\section{Observations and Data Reduction}

ASCA observed DEM L316 on 27 August, 1999.  X-rays were focused by the
four XRTs (X-ray Telescopes) separately onto two GIS (Gas Imaging
Spectrometer) and two SIS (Solid-state Imaging Spectrometer)
detectors.  The instruments are sensitive in the energy bands
 0.7--10 keV (GIS) and
0.4--10 keV (SIS).  The GIS and SIS were operated in PH mode and 1-CCD
faint mode, respectively. The roll angle of the observaion was
adjusted so that at the position of DEM L316 the stray light from LMC
X-1 was blocked by the mirror support housing.  Details of the ASCA
telescope and detectors are presented in Serlemitsos et al.~(1995),
Ohashi et al.~(1996) and Burke et al.~(1994).

The X-ray data were first screened with the usual criteria: data
obtained during South Atlantic Anomaly passages, with a cut-off
rigidity lower than 4 GeV c$^{-1}$, or with an elevation angle to the
limb of the Earth less than 5$^{\circ}$ were rejected. 
For the GIS, rise-time information was used to
further reject non-X-ray events.  For the SIS, we also rejected events
obtained when the elevation angle to the limb of the bright Earth was
smaller than 25$^{\circ}$ and with a cut-off regidity lower than
4GeV c$^{-1}$.
Hot and flickering pixels were removed and only standard event grades 0, 2, 3,
and 4 were used. 
 The net exposure time after screening was 6.0
$\times$10$^{4}$ s (SIS) and 5.6$\times$10$^{4}$ s (GIS).  
The increase in the charge-transfer-inefficiency of the SIS
CCDs over the course of the ASCA mission has resulted in a number of
deleterious effects including degraded spectral resolution and a
decrease in the detection efficiency for X-ray photons (due to grade
migration and on-board event rejection).  We therefore applied the RDD
(Residual Dark Distribution) correction as described by Dotani et
al.~(1997) to recover some of the lost spectral resolution.  This
results in an energy resolution for the SIS of $\sim$7\% (FWHM) at 1
keV.  The decrease in X-ray photon detection efficiency is incorporated
into the response function of the SIS.

We also extracted spectra from ROSAT PSPC (Position Sensitive
Proportional Counter) archival data (sequence number 500259)
originally observed in October 1993 for 4.0 ks.  The PSPC was
sensitive over the energy range 0.1--2.4 keV and had an energy
resolution at 1 keV of $\sim$43\% (FWHM).  The on-axis angular
resolution of the PSPC was 13$^{\prime\prime}$ (50\% encircled energy
radius) (Pfeffermann et al. 1987), 
which was sufficient to cleanly separate the X-ray emission
from the two shells in DEM L316.

\section{Analyses and Results}

\subsection{ SIS and PSPC X-ray Spectra}

In Figure 1 we show the ASCA SIS image in two energy bands; the soft
(0.4--1.2 keV) and hard (1.2--7.0 keV) X-ray band images are shown as
the grayscale and contour map, respectively.  We see X-ray emission
from both radio shells A and B. Since the half-power diameter of the
ASCA XRT ($\sim$3$^\prime$) is comparable to the angular separation
between the shells ($\sim$2.5$^\prime$), the SIS data do not cleanly
resolve the two SNRs and therefore each shell is contaminated by
X-rays that spill-over from the other shell.  In order to minimize
this spill-over effect and take advantage of the sharp core of the
ASCA XRT point-spread-function, we carefully selected the source and
background regions. The source regions, as shown in Figure 2a, were
rather small ellipses with major axes of 1$^\prime$ centered on the
peak emission from each shell. The background regions were the same
size as the source regions, were situated beyond the remnant, and were
located so that the source and background regions for one shell were
roughly the same distance from the source region of the other shell.
We note that the overall ASCA SIS fluxes are not well determined since
only a portion of each shell was included in the small spectral
extraction regions used.  In order to obtain a more accurate estimate
of the overall flux of each shell and to provide stronger constraints
on the soft X-ray absorption toward the SNR, we thus extracted ROSAT
PSPC spectra using the source and background regions shown in Figure
2b.  Clearly the emission from shells A and B is cleanly separated in
the PSPC data.

The SIS and PSPC spectra were simultaneously fitted to an
optically-thin thermal plasma spectral model.  All spectral parameters
were linked between the data sets for a given shell with the exception
of the SIS and PSPC normalization factors which were each allowed to
vary freely. Below we present results based on the nonequilibrium
ionization (NEI) spectral code by Masai (1994) (hereafter refered to
as the Masai model) although, as a cross check, the NEI model of
Hughes \& Singh (1994) was also used to verify results.  In general
there was good agreement between the two NEI spectral calculations and
in particular we find roughly the same best-fit elemental abundances.
The specific NEI model employed was the single-timescale,
single-temperature model, which parameterizes the NEI condition
through a quantity called the ionization timescale ($n_{\rm e}t_{\rm
i}$). Here $n_{\rm e}$ and $t_{\rm i}$ are the mean electron density
and elapsed time after the plasma was heated to the temperature $kT$.
The equilibrium ionization condition is attained for values of $n_{\rm
e}t_{\rm i} > 10^{12}\,\rm cm^{-3} s$.  In the following we refer to
this as collisional ionization equilibrium (CIE).

\subsubsection{Shell A}

In our NEI fits to the data of shell A, we first allowed the
temperature, column density and ionization parameter to be free, while
the abundances of all the elements were fixed to be 0.3 times solar,
the average in the LMC (see Russell \& Dopita 1992, Hughes et
al.~1998).  This model, however, was statistically unacceptable (the
$\chi^2$ of 85.4 for 52 degrees of freedom can be rejected at the
99.7\% confidence level), and furthermore the data showed an excess of
emission relative to the model around 1 keV. This energy band contains
the L-shell lines of iron and the K$\alpha$ lines of neon, which
raises the interesting possibility that one or both of these species
may be relatively overabundant in shell A.

In order to test this hypothesis we tried several model fits including
cases with (1) only the neon abundance free, (2) only the iron
abundance free, and (3) both the iron and neon abundances free.  In
these fits the column density to shell A was fixed to $N_{\rm H} = 3.8
\times 10^{21}\,\rm cm^{-2}$, a value that is consistent with our
preliminary fits (described in the preceding paragraph) and an
independent estimate from the HI column to the SNR (Williams et
al.~1997).  The abundances of the other elemental species were fixed
to be 0.3 times solar, as above.  Model 1 was rejected since it was
unable to account for the excess residual emission between 1 to 2 keV,
while Models 2 and 3 were both acceptable from a statistical point of
view.  In the two latter cases, the iron abundance was found to be
significantly higher than the LMC mean value.  However, a similar
statement about the abundance of neon in shell A could not be made,
since the neon abundance determined in model 3 was not well
constrained. 
In table 1 we present the best-fit parameters from our
case 2 model fits.  Figure 3a shows the spectral data, best-fit
model spectrum and residuals. The best-fit value of the iron abundance
is $1.9_{-0.6}^{+0.9}$ times solar, a factor of 5 more than the mean
LMC value, which clearly indicates a large overabundance of iron in
shell A. We also note that the ionization timescale range allows for
the possibility that the X-ray--emitting plasma is in CIE, as shown in
the two-dimensional $\chi^2$ contours of $kT$ vs.\ $n_{\rm e}t_{\rm
i}$ (Figure 4a).

\subsubsection{Shell B}

For this shell the NEI model with fixed LMC mean abundances (0.3
solar) provided an acceptable fit to the spectrum.
As before the best-fit column was fixed to the
value $N_{\rm H} = 3.8 \times 10^{21}\,\rm cm^{-2}$.    
The $\chi^2$ confidence contours between ionization timescale and 
temperature are shown in Figure 4b.
This plot indicates that we could not constrain whether the plasma in 
shell B is in NEI or CIE condition because the confidence level for 
rejecting the CIE model is slightly low (68\%), though the best fit 
ionization timescale favors the NEI case.
Therefore, we also carried out the spectral fitting for the CIE case.
The best fit parameters for both the NEI and CIE case are in Table 1.
Figure 3b shows plots of the data, model, and residuals only 
for the NEI case, 
because the best-fit spectrum in CIE is almost the same as that in NEI.

\subsubsection{X-ray spectra with GIS}

The spatial resolution of the GIS is somewhat poorer than that of the
SIS, which makes separating the X-ray emission from the two shells in
DEM L316 nearly impossible.  Thus we made a composite GIS spectrum
containing the total emission from shells A and B again summing data
from both detectors.  We fit this GIS spectrum using the the best fit
SIS/PSPC spectral parameters and found the fit to be acceptable,
thereby establishing consistency between the GIS and SIS/PSPC data.

\section{Discussion}
The main difference between the spectral characteristics of shell A
and B is the strong emission line structure at about 1 keV seen in the
spectrum of shell A.  Our spectral analysis of shell A suggests an
enhanced iron abundance.  
It is well known that Type Ia SNe are
efficient producers of iron (Nomoto, Thielemann, \& Yokoi 1984), while
massive-star core-collapse SNe tend to produce proportionally more of
the lower atomic number species, such as oxygen, neon, and magnesium
(Tsujimoto et al.~1995).
Therefore one might argue that the emission line structure around 1 keV 
may be described by a combination of K$\alpha$ lines from 
oxygen and neon, which are major elements yielded by type II SNe, 
rather than L shell lines from iron. 
To check this possibility, we further treated the abundances of neon, oxygen, and 
iron as free parameters and fitted the shell A spectra to the NEI model. 
Then we found that the abundance of iron was again  significantly 
higher than the mean LMC value. However, a similar statement 
about the abundances of oxygen and neon could not be made, since 
the abundances of both of these elements were not well constrained.
At least one other middle-aged LMC SNR (DEM
71) shows evidence for enhanced iron in its global integrated ASCA
spectrum (Hughes et al.~1998).  
We, therefore, believe that the enhanced iron 
in shell A is strong evidence in support of the origin of shell A as
a Type Ia SN.

Assuming that the electron and ion densities in the plasma are identical,
that the X-ray emitting plasma has uniform density 
and that the distance to DEM L316 ($D$) is 50
kpc, we can estimate the ion density ($n_{\rm i}$) in both shells
through 
\begin{equation}
\label{eq:Laplace}
n_{\rm i} = 1.7 N_{12}^{0.5} (V/10^{59}{\rm cm}^{3})^{-0.5} (D/50 {\rm kpc}) {\rm cm}^{-3},
\end{equation}
where $N_{12}$ is the emission measure derived from the spectral fits
(in units of $10^{12}\rm\, cm^{-5}$) and $V$ is the volume of the
X-ray emitting plasma (in units of cm$^3$). 

We estimate the volume using the ROSAT HRI morphology (Williams et al.~1997) 
as follows.  
For  shell A, we assume that the plasma fills a spherical 
region with a radius of 11 pc.
Because shell B has limb-brightened morphology,
we tentatively assumed that the Sedov model applies to shell B, 
i.e.,
the thickness of shell B is one twelfth of the shell radius 
(shell radius was assumed to be 15 pc, Williams et al.~1997).

From equation (1), we obtain values of $\sim$0.4 cm$^{-3}$
 for the ion density of the plasma in shell A.
Adopting this density and the best-fit ionization parameter 
$n_{\rm e}$$t_{\rm i}$, we derive an ionization timescale age for shell A 
of $\sim$3.9$\times$10$^4$ yr.

For shell B, the spectral fitting gives two local minimums for the NEI 
conditions of $n_{\rm e}$$t_{\rm i}$ $\sim$ 10$^{10.7}$ (case 1) and 
$n_{\rm e}$$t_{\rm i}$ $\sim$ 10$^{12}$ (case 2).  
Depending on these two cases, we obtain different physical parameters,
particularly for the ionization age.
In case 1, the plasma density is $\sim$ 0.5, for
which case the ionization parameter ($n_{\rm e}$$t_{\rm i}$) gives an age 
($t_{\rm i}$) of $\sim$3.2$\times$10$^3$ yr.
However, in case 2, the plasma density is $\sim$ 0.8 and the age is 
estimated to be $>$ 4.2$\times$10$^4$ yr  
(the CIE condition is attained for values of  
  $n_{\rm e}$$t_{\rm i}$ $>$ 10$^{12}$).  
Shell B is larger and has lower surface
brightness than shell A, which suggest that shell B is the older of the
two.  We thus infer that the evidence favors case 2, which places shell B
at an older age ($>$ 4.2$\times$10$^4$ yr) than that found for shell A above.

Finally we propose, from the totally different spectral shapes of shell 
A and B,  
that these two shells were not due to a single supernova explosion into a
pre-existing bubble made by a stellar wind or an SN.
We also propose, from the iron over abundance seen in the spectrum, 
that the origin of shell A is a Type Ia.

\section{Summary}
The first detailed spectral analysis of DEM L316 was demonstrated with the 
 ASCA satellite. The spectra of shell A exhibited strong L-shell 
lines of iron, 
thus we conclude that shell A has Type Ia SN origin.
On the other hand, the shell B spectra was well reproduced with the
 thin-thermal plasma model whose abundance was similar to that of the
 interstellar medium of the LMC. These totally different spectral 
characteristics strongly contradict the hypothesis that two shells were 
due to a single supernova explosion into a pre-existing bubble made by a 
stellar wind or an SN.
We could not obtain the evidence which 
positively supports the collision between two SNRs nor could we 
get a strong constraint for the shell B progenitor.
To further investigate these problems more spatially  sensitive observations 
 in X-ray band will be needed, such as with the Chandra satellite.

\par
\vspace{1pc}\par
JPH extends his thanks to Monique Arnaud for her support and
hospitality during the course of this project.  Partial funding was
also provided by NASA grant NAG5-6420 to Rutgers University.  We thank
Marc Gagn\'e for his contributions to the initial planning of these
observations. This research has made use of data obtained from the
High Energy Astrophysics Science Archive Research Center (HEASARC),
provided by NASA's Goddard Space Flight Center.
MN and JY are financially supported by JSPS Research Fellowship 
for Young Scientists.

\begin{table*}[t]
 \begin{center}
Table.~1\hspace{4pt}Fitting results.
 \end{center}
  \begin{tabular*}{\textwidth}{@{\hspace{\tabcolsep}
\extracolsep{\fill}}p{12pc}ccc} \hline\hline\\[-6pt]
 & Shell A & Shell B (CIE) & shell B (NEI)\\
\hline
$N_{\rm H}$ ($10^{21}\rm\, cm^{-2}$)
  & 3.8$^\dagger$
  & 3.8$^\dagger$
  & 3.8$^\dagger$ \\
$kT_{\rm e}$ (keV)
  & 0.90 (0.86--1.09) 
  & 0.72 (0.68--0.76) 
  & 1.13 (0.87--1.52) \\
log~$n_{\rm e}t_{\rm i}$ (cm$^{-3}$~s)
  & 11.74 (11.07--13.00) 
  & 13.00$^\dagger$     
  & 10.73 (10.53--11.07)\\
Fe $^\ast$  
  & 1.9 (1.3--2.8)  
  & 0.3$^\dagger$
  & 0.3$^\dagger$ \\ 
Other species $^\natural$    
  & 0.3$^\dagger$  
  & 0.3$^\dagger$
  & 0.3$^\dagger$   \\
EM$^\ddagger$ (10$^{10}$ cm$^{-5}$)
   & 6.9 (3.7--7.0)
   & 16.6 (14.6--18.6)
   & 8.4 (7.0--10.1) \\
$\chi^2$ (d.o.f) 
   & 50.9 (52) 
   & 46.1 (45)
   & 43.4 (44) \\
\hline
\end{tabular*}
 \vspace{6pt}\par\noindent
\par\noindent
 Parentheses indicate 90\% confidence intervals for a single
 relevant parameter ($\Delta\chi^2 = 1$)
\par\noindent
$^\dagger$ Fixed parameter
\par\noindent
$^{\ast}$ Abundance of iron relative to the solar value.
\par\noindent
$^{\natural}$ Abundances of other elements relative to their solar values.
\par\noindent
$^{\ddagger}$ Emission measure, defined as: ($n_{\rm i}^2~V$)/(4 $\pi$ $D^2$). 
See text for details.
\par\noindent
\end{table*}

\small
 \section*{References}
\re Burke B.E., Mountain R.W., Harrison D.C., Bautz M.W., Doty J.P., 
    Ricker G.R., Daniels P.J.\ 1991, IEEE Trans.\ ED-38, 1069
\re Dotani T., Yamashita A., Ezuka H., Takahashi K., Crew G., Mukai K., 
    the SIS team\ 1997, ASCA news 5, 14
\re Feast M.W.\ 1991, in IAU Symp. 148, The Magellanic Clouds, ed. R.Haynes \& D.Milne 
(Dordrecht : Kluwer), 1
\re Hughes J.P., Hayashi I., Koyama K.\ 1998, ApJ 505, 732
\re Hughes J.P., Singh K.P.\ 1994, ApJ 422, 126
\re Lasker B.M.\ 1981, PASP 93, 422
\re Mathewson D.S., Clarke J.N.\ 1973, ApJ 180, 725
\re Mills B.Y., Turtle A.J., Little A.G., Durdin J.M.\ 1984, Australian J.\ 
 Phys.~37, 321
\re Nomoto, K., Thielemann, F.-K., Yokoi, M.\ 1984, ApJ 286, 644
\re Ohashi T., Ebisawa K., Fukazawa Y., Hiyoshi K., Horii M., Ikebe Y., 
    Ikeda H., Inoue H. et al.\ 1996, PASJ 48, 157
\re Pfeffermann E., Briel U.G., Hippmann H., Kettenring G., Metzer G., Predehl P., Reger G., Stephan K.H. et al. \ 1987, Proc. SPIE 733,519
\re Russell S.C., Dopita M.A.\ 1992, ApJ 384, 508
\re Serlemitsos P.J., Jalota L., Soong Y., Kunieda H., Tawara Y., Tsusaka Y., 
    Suzuki H., Sakima Y. et al. \ 1995, PASJ 47, 105
\re Tsujimoto T., Nomoto K., Yoshii Y., Hashimoto M., Yanagida S., Thielemann F.K. \ 1995, MNRAS 277, 945
\re van den Bergh, S.\ 2000, PASP 112, 529
\re Williams R.M., Chu Y.H., Dickel J.R., Beyer R., Petre R., Smith R.C.,
    Milne D.K.\ 1997, ApJ 480, 618

\newpage
\begin{figure}[th]
\hspace*{8mm} \psbox[xsize=0.4\textwidth]{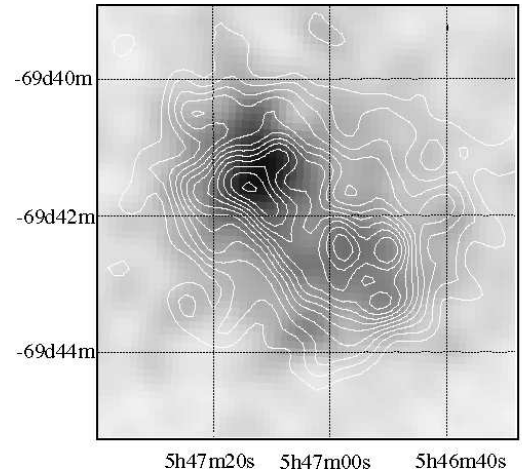}
\caption{ASCA SIS image of DEM L316 (SNR 0547$-$69.7) after summing both
detectors and smoothing with a gaussian filter of $\sigma =
15^{\prime\prime}$.  The grayscale and contours represent the 0.4--1.2
keV and 1.2--7.0 keV bands, respectively.  Contours are linearly
spaced in steps of $2.8\times10^{-4}$ cts~s$^{-1}$~arcmin$^{-2}$ and
sky coordinates are epoch J2000.  The two X-ray sources correspond to
shell A (northeastern source) and shell B (southwestern source),
respectively.}
\end{figure}


\begin{figure}[th]
\hspace*{8mm} \psbox[xsize=0.4\textwidth]{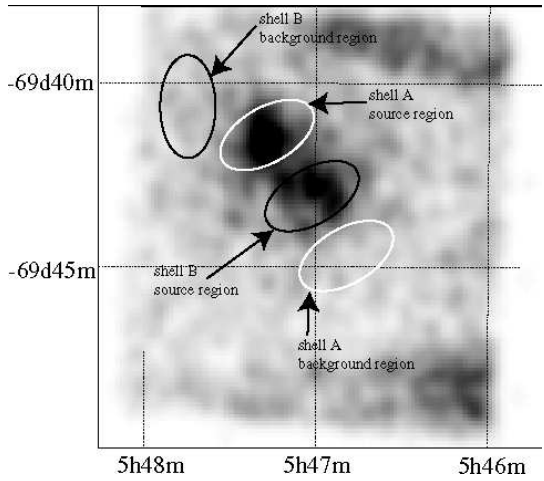}

\vspace*{3cm}
\hspace*{8mm} \psbox[xsize=0.4\textwidth]{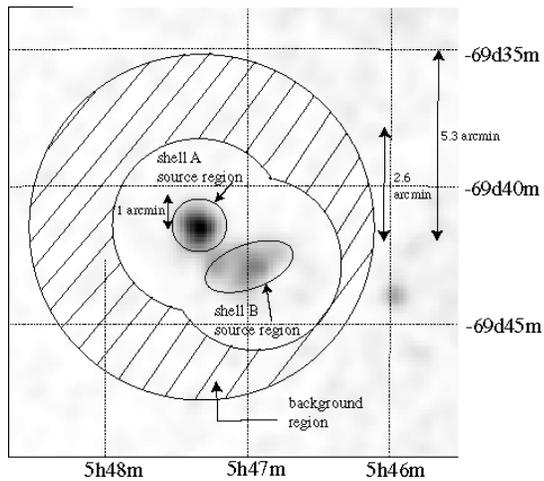}
\caption{The source and background regions (see text) overlaid on the 
ASCA SIS (a) and ROSAT PSPC (b) total band images.}
\end{figure}

\begin{figure}[th]
\hspace*{8mm} \psbox[xsize=0.4\textwidth]{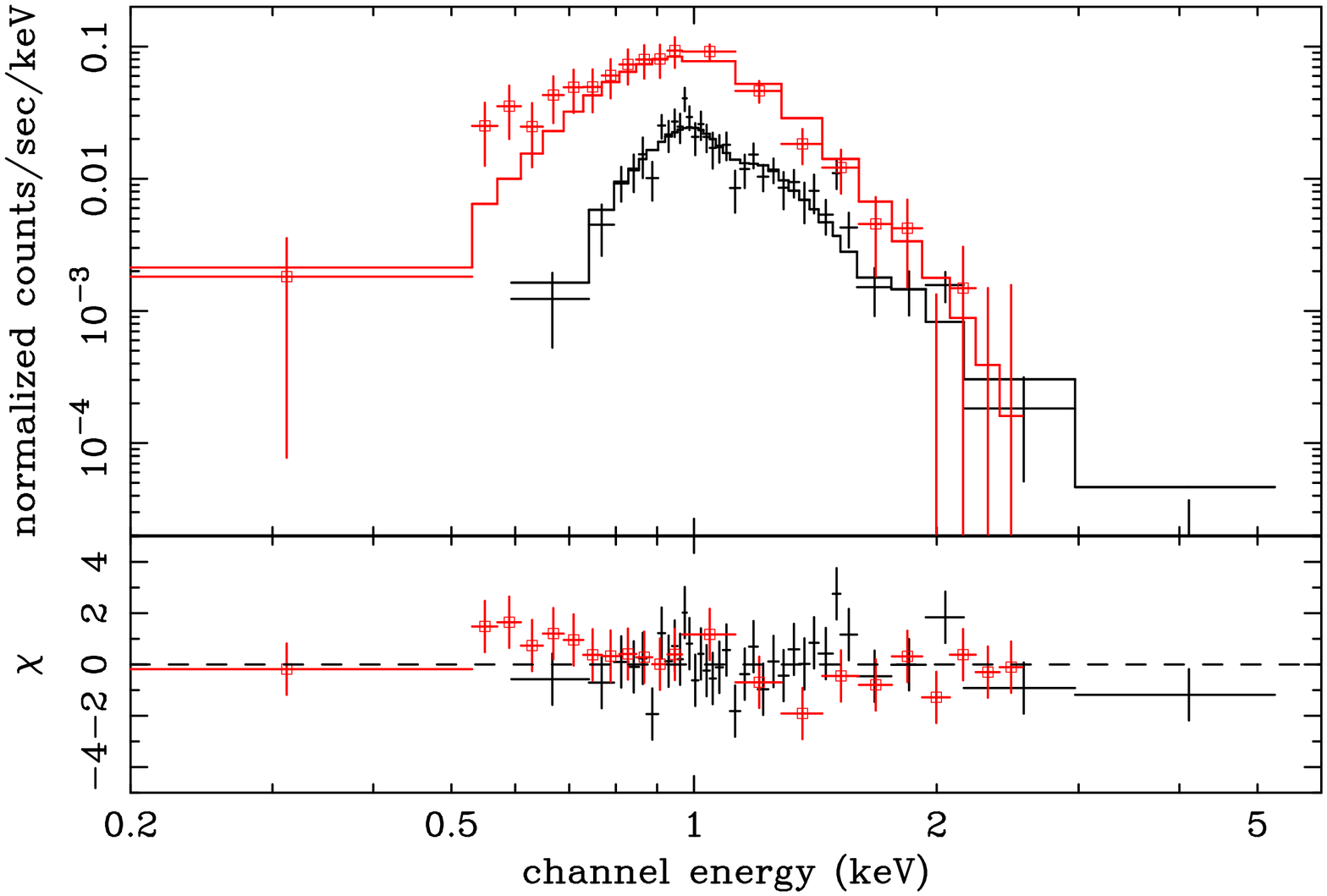}

\hspace*{8mm} \psbox[xsize=0.4\textwidth]{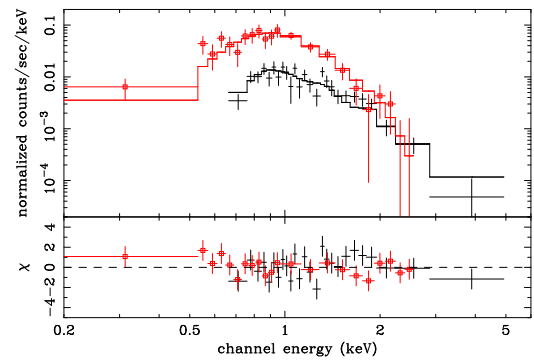}
\caption{X-ray spectra of shell A (a) and shell B (b) from the ASCA SIS
(crosses) and ROSAT PSPC (boxes).  The top panels show the data and
best-fit models, while the lower panels shown the residuals.  Errors
are purely statistical at 1 $\sigma$ confidence.  The spectra appear
to differ the most around 1 keV.  We attribute the more sharply peaked
emission from shell A near 1 keV to enhanced iron L line emission (see
text for details).}
\end{figure}

\begin{figure}[th]
\hspace*{8mm} \psbox[xsize=0.4\textwidth]{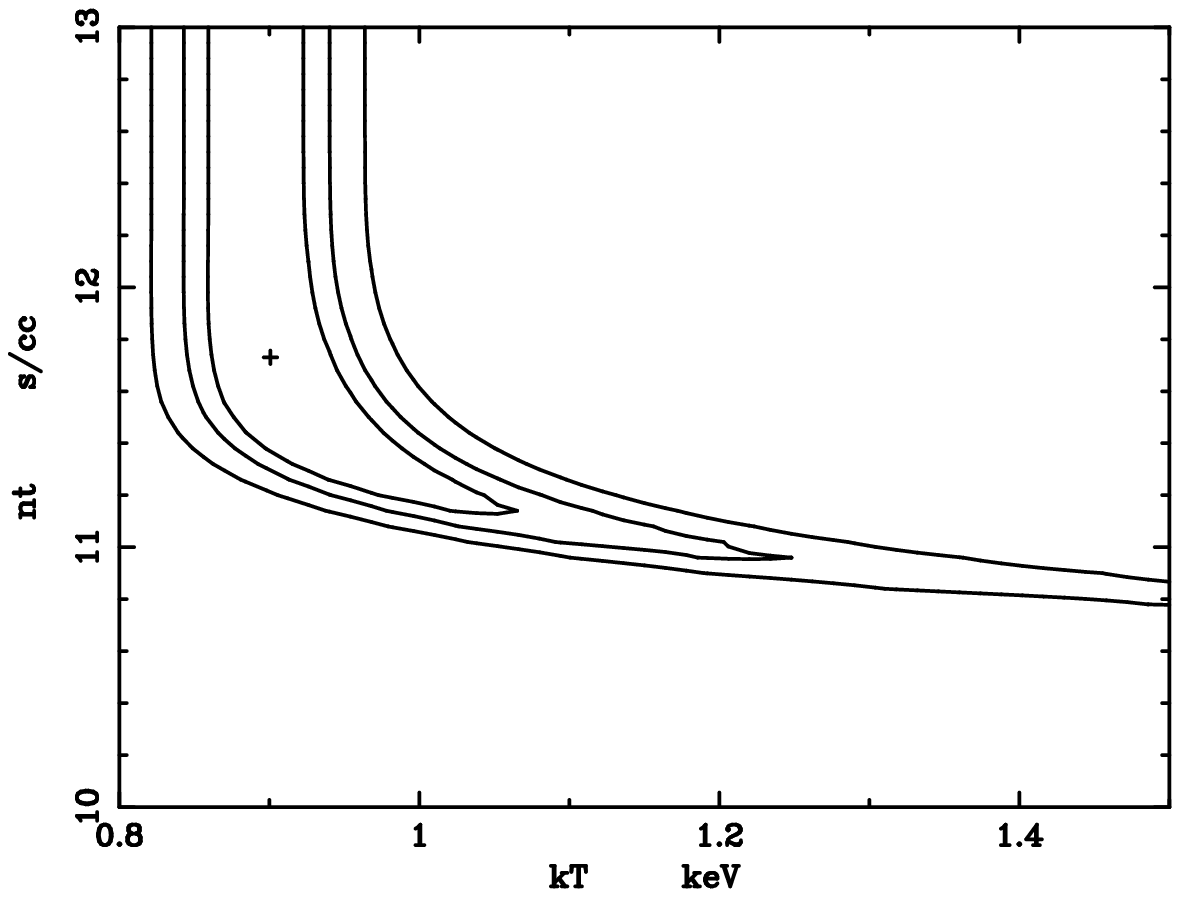}

\hspace*{8mm} \psbox[xsize=0.4\textwidth]{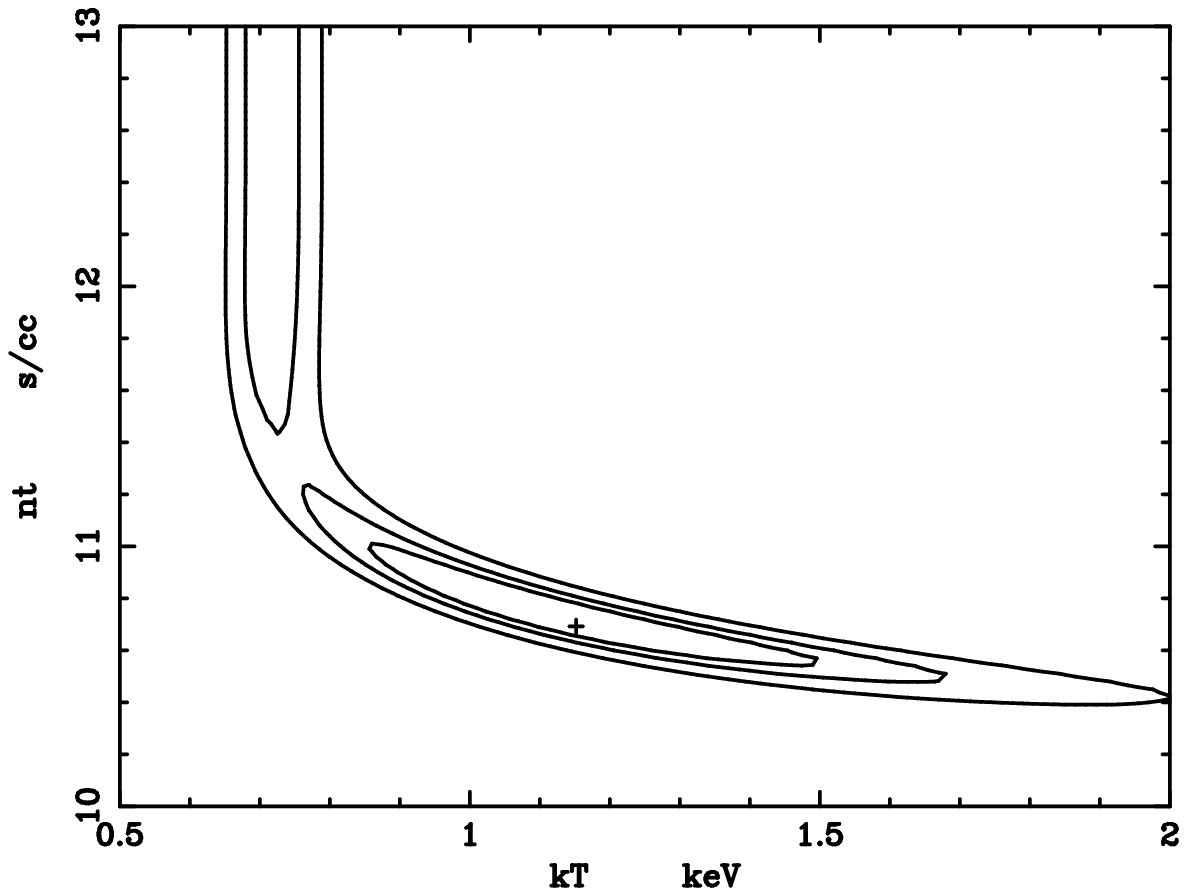}
\caption{Confidence contours between fitted electron temperature ($kT$) and
ionization timescale ($nt$) for shell A (a) and shell B (b).  Contours
are shown at the 68\%, 90\% and 99\% confidence levels.  The best-fit
parameters are indicated by crosses.}
\end{figure}

\end{document}